\documentclass[%
 reprint,
 amsmath,amssymb,
 aps,
]{revtex4-1}

\usepackage{graphicx}
\usepackage{dcolumn}
\usepackage{bm}

\newcommand{\q}{\quad}
\newcommand{\qq}{\qquad}
\newcommand{\ket}[1]{\left|{#1}\right\rangle}
\newcommand{\bra}[1]{\left\langle{#1}\right|}
\newcommand{\aver}[1]{\left\langle{#1}\right\rangle}

\usepackage{hyperref}
\usepackage{caption} 
\usepackage{booktabs,siunitx}
\usepackage{enumerate}
\usepackage{float}

\begin{document}


\title{Bifurcations, time-series analysis of observables, and network properties in a tripartite quantum system}

\author{Pradip Laha}
 \altaffiliation[]{pradip@physics.iitm.ac.in}
\author{S. Lakshmibala}%
\affiliation{%
 Department of Physics, IIT Madras, Chennai 600036, India\\}%
\author{V. Balakrishnan}%
\affiliation{Department of Physics, IIT Madras, Chennai 600036, India\\}%




\date{\today}

\begin{abstract}
 In a tripartite system comprising a $\Lambda$-atom  interacting with two radiation fields in the presence of field nonlinearities and an intensity-dependent field-atom coupling, striking features have been shown to occur  in the dynamics of the mean photon number  $\aver{N_{i}(t)}$ ($i = 1,\,2$) corresponding to either field.   In this Letter,  we carry out a detailed  time-series analysis and establish an  interesting   correlation  between the  short-time and long-time dynamics of $\aver{N_{i}(t)}$.   Lyapunov exponents, return maps, recurrence plots, recurrence-time  statistics, as well as  the clustering coefficient and the transitivity of networks constructed from the time series,   are studied as functions  of the intensity parameter $\kappa$. These are  shown to  carry signatures of a   special value $\kappa = \overline{\kappa}$.     Our work also exhibits how  techniques from nonlinear dynamics help analyze the behavior of observables in multipartite quantum systems.

\begin{description}
\item[PACS numbers]
05.45.Tp; 42.50.-p; 05.45.-a
\end{description}
\end{abstract}

\pacs{Valid PACS appear here}
\maketitle


\section{\label{sec:level1} Introduction}
   In machine-learning protocols and in  solutions to problems involving computationally intensive procedures,  reduction of a large data set to a considerably smaller  optimal set~\cite{seth_lloyd}  poses a challenge. Network analysis provides  an effective solution to this problem, particularly  in the context of  machine learning~\cite{seth_lloyd,rebentrost}.  Further, long time series of classical dynamical variables need to be investigated in a variety of  contexts  ranging from weather forecasting and climate analysis to medical research~ \cite{zou,gao1,donges,marwan2,ramirez}.  The tools of network analysis have been used in these examples  to select  an optimal subset of data points in the time series for  investigating  the underlying dynamics.  Different indicators that capture the important features of the time series have been identified  from such smaller data sets in model classical systems~\cite{kurths_phys_rep}. 
  
 In the quantum context, too,  large data sets pertaining to the dynamics of appropriate observables 
  have  been either  measured  in specific experiments, or obtained through numerical simulations in  certain cases (see, e.g.,~\cite{wittek}). Quantum optics provides a wide range  of experimentally realizable examples in which both the short-time behavior and the long-time behavior exhibit interesting dynamical properties. The diversity of the ergodicity properties displayed by experimentally measurable quantum mechanical expectation values  provides the motivation to employ the  tools of time-series analysis such as first-return-time distributions, recurrence plots, calculation of Lyapunov exponents, and so on, to understand the temporal behavior  of these observables. In the case  of the bipartite  model considered in Ref.~\cite{puri}, investigations along these lines have been carried out in earlier work~\cite{sudh_pla,sudh_epl}. This model describes a multilevel atom of a nonlinear medium interacting with a radiation field. For sufficiently  high nonlinearity (relative  to the interaction strength),  it has been shown that the dynamics of the mean photon number can exhibit exponential instability, as indicated by a positive maximal Lyapunov exponent (MLE). In the case of a multipartite model of a three-level  $\Lambda$-atom interacting with either one or two radiation fields, a detailed time-series analysis~\cite{athreya} of the mean photon number of either field, $\aver{N_{i}(t)}$ ($i=1,\, 2$), reveals a wide range of ergodic behavior, depending sensitively  on the strength of the nonlinearity and the degree of coherence of the initial state  of the radiation field(s).

Another aspect of the dynamics of quantum observables pertains to the  nonclassical effects that are manifested  in the time series. These effects  include quadrature and entropic squeezing~\cite{laha2,laha3,sharmila1}, sudden death of the entanglement between subsystems~\cite{eberly}, and the collapse of the entanglement to a constant non-zero value over a significant time-interval~\cite{laha1,laha3}. This entanglement collapse is highly sensitive to the initial state considered and  the field nonlinearities. In particular, in the tripartite model mentioned above, the entanglement of the atomic subsystem collapses  to a fixed non-zero value over a significant time interval~\cite{laha1}. This  is also reflected in the dynamics of both the mean and the variance of the photon number corresponding to either of the two field modes. 
 
 Apart from field nonlinearities, several experimental techniques have been developed in recent years to create nonlinear interactions between  photons and single atoms (see, for instance,~\cite{chin_nc}). The rather simple procedure of coupling light to a single atom by focusing the incident photons onto the atom with a lens only provides moderate interaction strengths. This hurdle has been overcome recently by adapting a super-resolution imaging technique, namely, 4Pi microscopy.  A test of the nonlinear interaction is the experimental observation of modified photon statistics of the transmitted field~\cite{chin_nc}.

  If the field-atom coupling in the above model describing a $\Lambda$-atom interacting with light~\cite{laha1} is dependent on the field intensity,  new phenomena occur. 
  Several forms of intensity-dependent couplings (IDCs)  have been introduced and analyzed in the literature to understand the dynamical behavior of the system. Some of them are:  $f(N_{i}) = N_{i}^{1/2}$~\cite{buck}, 
  $N_{i}^{-1/2}$~\cite{sudarshan}, $(1+\kappa\,N_{i})^{1/2}$~\cite{siva} and $(1 - \tfrac{1}{2}\eta^{2}\, N_{i})$~\cite{barzanjeh}.  Here $\kappa$ is the `intensity parameter' and  $\eta$  is another such parameter (the Lamb-Dicke parameter). In particular, for an intensity-dependent coupling of the form $f(N_{i}) = (1 + \kappa\, N_{i})^{1/2}$,  the mean photon number $\langle N_{i}\rangle$ exhibits an interesting  bifurcation cascade as $\kappa$ is varied from 0~to~1~\cite{laha1}. 
  Significantly,  there is an underlying algebra of field operators associated with this particular functional form of the coupling.  Let  $R  =  a\, f(N)$ and $R_{0} = \tfrac{1}{2}+\kappa(N+\tfrac{1}{2})$ where 
 $N = a^{\dagger}a$  ($a$ and $a^{\dagger}$ are the photon annihilation and creation operators, respectively) 
and $f(N) =  (1 + \kappa \,N)^{1/2}$.  
  Then the operators  $R, R^{\dagger}$ and  $R_{0}$ satisfy  a closed algebra under commutation:    
\begin{equation}
  [R, R^{\dagger}] = 2R_{0}, \q[R, R_{0}] = \kappa R, \q [R^{\dagger}, R_{0}] = -\kappa R^{\dagger}.
\end{equation}
This is a deformation of the Lie algebra corresponding 
to the group SU(1,1), with $\kappa$ as the deformation 
parameter. The limiting case  $\kappa = 0$ 
reduces to the usual 
Heisenberg-Weyl algebra for the field operators, while the  case  $\kappa = 1$  yields the Lie algebra of the group 
SU(1, 1).  Hence, by varying $\kappa$ from 
$0$ to $1$, we can examine the effects 
of a continuous change in the algebra of the operators relevant to the 
system on the 
dynamics (specifically, on the ergodic properties of a suitable dynamical variable), although this precise 
form of the IDC has not been realized experimentally as yet.  It has also been verified that, for the IDCs  
$f(N_{i}) = N_{i}^{1/2}$ and $N_{i}^{-1/2}$, 
the dynamics is not as rich as it is for  
$f(N_{i}) = (1+ \kappa\, N_{i})^{1/2}$~\cite{laha1}. 
The possibility  
of  experimentally  exploring   the consequences of the specific form $f(N_{i}) = (1 - \tfrac{1}{2}\eta^{2}\, N_{i})$  has been examined~\cite{barzanjeh}, although in the system under consideration this IDC does not 
lead to  the diverse dynamical effects 
displayed when $f(N_{i}) = (1 + \kappa\, N_{i})^{1/2}$. 
For all the reasons described in the foregoing, 
 we focus on the latter choice of IDC in what follows.

Recently,  an $\epsilon$-recurrence network has been constructed~\cite{laha4} for the time series of $\langle N_{i}(t)\rangle$. It has been shown that the signatures of the bifurcation cascade  are also captured in the manner in which the clustering coefficient and the transitivity of the network vary with $\kappa$. In the light of this fact, certain significant questions arise: do these network quantifiers reflect the properties of the cascade if the $\epsilon$-recurrence network is replaced by other types of networks?  Do network quantifiers such as the transitivity, clustering coefficients, etc., also exhibit interesting features deduced from  the MLEs, return maps, recurrence plots and recurrence-time distributions? In particular, is there a relation between the manner in which these important elements in the theory of dynamical systems (on the one hand) and  network properties (on the other) vary with changes in the relevant parameters of the system (in this case, $\kappa$)? The purpose of this Letter is to address these questions.
  
The relevant features of the  tripartite model considered are given in  Sec.  \ref{sec:model}. In Sec. \ref{sec:bifurcation}, we present a brief review of the manner in which the  properties of the network correlate with the bifurcation sequence exhibited by~$\aver{N_{i}}$. Section \ref{sec:time_series} is devoted to  the MLEs, return maps, recurrence plots and first-return-time distributions for  different values of $\kappa$. The occurrence and effect of a special value  $\bar{\kappa}$ on the network properties such as the clustering coefficients and the transitivity  are also discussed.  The inferences to be drawn from this work are summed up in Sec. \ref{sec:conclusion}.  The key steps in the derivation of the state of the system at any  time $t$ are outlined in the Appendix. 

\section{\label{sec:model} The tripartite model}

We consider a  $\Lambda$-atom inside a cavity, interacting with a probe field $F_{1}$ and a coupling field $F_{2}$ with respective frequencies $\Omega_{1}$ and $\Omega_{2}$ (see, e.g., Fig. 1 of Ref.~\cite{laha4}). The corresponding photon annihilation and creation operators are $a_{i}$ and  $a^{\dagger}_{i}$, $(i = 1,\,2)$.  The two lower energy states of the $\Lambda$-atom are denoted by $\ket{1}$ and $\ket{2}$, respectively, and the excited state by $\ket{3}$. The fields $F_{1}$ and $F_{2}$ induce $\ket{1}\leftrightarrow \ket{3}$ and $\ket{2}\leftrightarrow \ket{3}$  transitions, respectively. The transition $\ket{1}\leftrightarrow \ket{2}$ is dipole-forbidden. The Hamiltonian incorporating field nonlinearities and field-atom IDC is given 
(with $\hbar = 1$) by
\begin{align}
H  =  \sum\limits_{j=1}^{3} \omega_{j} \sigma_{jj}  + \sum\limits_{i=1}^{2}&\Big\{ \Omega_{i} \,a_{i}^{\dagger} a_{i}  +  \chi \,a_{i}^{\dagger 2} a_{i}^{2} \nonumber \\
                                        &\hspace{-4.5ex}+  \lambda \big[a_{i} \, f(N_{i}) \,\sigma_{3i} +  f(N_{i})\, a^{\dagger}_{i} \,\sigma_{i3})\big] \Big\}.
\label{eqn:lambda_hamiltonian}
\end{align}
Here, $\sigma_{jk} \equiv \ket{j}\bra{k}$ are the Pauli spin operators  
(where $\ket{j}$ denotes  an atomic state), and $\omega_{j}$ are positive constants.  $\chi$  is the strength of the nonlinearity in both the fields, and $\lambda$ is the atom-field coupling parameter.  As mentioned earlier, the IDC $f(N_{i}) = (1+\kappa_{i}\,N_{i})^{1/2}$  with $N_{i} = a^{\dagger}_{i} a_{i}$. For simplicity, in our calculations we set the detuning parameter to zero, i.e., $\omega_{3} - \omega_{i} - \Omega_{i} = 0\, (i=1,\, 2)$, and also set  $\kappa_{2} = 0$ and $\kappa_{1} = \kappa$ where  $0 \leqslant \kappa \leqslant 1$.

 The  $D_{1}$ transition in an ${^{85}}$Rb atom driven by a pump and a coupling field is an example of a $\Lambda$-atom interacting with two radiation fields.  In this system, the two ground states of the atom correspond to the hyperfine levels 5S$_{1/2}$ (F $=2$) and 5S$_{1/2}$ (F $= 3$), while the excited state corresponds to the 5P$_{1/2}$ in the $D_{1}$ line of the $^{85}$Rb atom. In a sense this is the `workhorse' of quantum optics. Several new phenomena including electromagnetically induced transparency (or absorption) have been 
 investigated extensively, both experimentally and theoretically,  in this system (for further details see, for instance, \cite{li_xiao,fleischhauer,yang_sheng,yang_xiao}). 

The field nonlinearity in our model is Kerr-like, i.e., of the form  $a_{i}^{\dagger 2} a_{i}^{2}$ (see Eq.\eqref{eqn:lambda_hamiltonian}). This type of nonlinearity  has been used in several theoretical models and is readily realizable experimentally. Exhaustive work on Kerr nonlinearities has been reported in the literature.  Some recent interesting results may be found in~\cite{honarasa_2012,ghasemian,wang_prl}. 

 We take  the initial states of both the fields to be the standard normalized coherent state (CS)  
 \begin{equation}
 \ket{\alpha} = \sum_{n=0}^{\infty} q_{n}(\alpha) \, \ket{n}, 
 \;\; q_{n}(\alpha)  = \frac{e^{-\vert\alpha\vert^{2}/2} \, 
 \alpha^{n}}{\sqrt{n!}},
 \label{csdefn}
 \end{equation}
 where  $\alpha \in \mathbb{C}$, and that of the  atom  to be the state $\ket{1}$.  Hence the initial state of the system is given by
\begin{equation}
 \ket{\psi (0)}  =  \sum\limits_{n=0}^{\infty}\sum\limits_{m=0}^{\infty} q_{n}(\alpha)  q_{m}(\alpha)  \ket{1; n; m},
\label{eqn:two_mode_lambda_initial_state}
\end{equation}  
where the labels in the ket vector refer respectively to the state of the atom and the number states of the two fields. Solving the Schr\"{o}dinger equation in the interaction picture, the state of the system at any time $t> 0$ is found to be of the form
\begin{align}
 \ket{\psi (t)} =  \sum\limits_{n=0}^{\infty}\sum\limits_{m=0}^{\infty} 
 &q_{n}(\alpha)  q_{m}(\alpha)  \Big\{ A_{nm}(t)  \ket{1; n; m}   \nonumber \\
      &+ B_{nm}(t) \ket{2; n-1; m+1} \nonumber \\
      & + C_{nm}(t) \ket{3; n-1; m} \Big\}.
\label{eqn:two_mode_lambda_interaction_state}
\end{align}
Explicit expressions are given in the Appendix for the 
time-dependent coefficients $A_{nm}, \,B_{nm},\,C_{nm}$, and for the reduced density matrices of the subsystems. 

\section{\label{sec:bifurcation} Bifurcation cascade and network analysis}
\begin{figure}
 \centering
 \includegraphics[height=4.5cm, width=8.35cm,angle=-00]{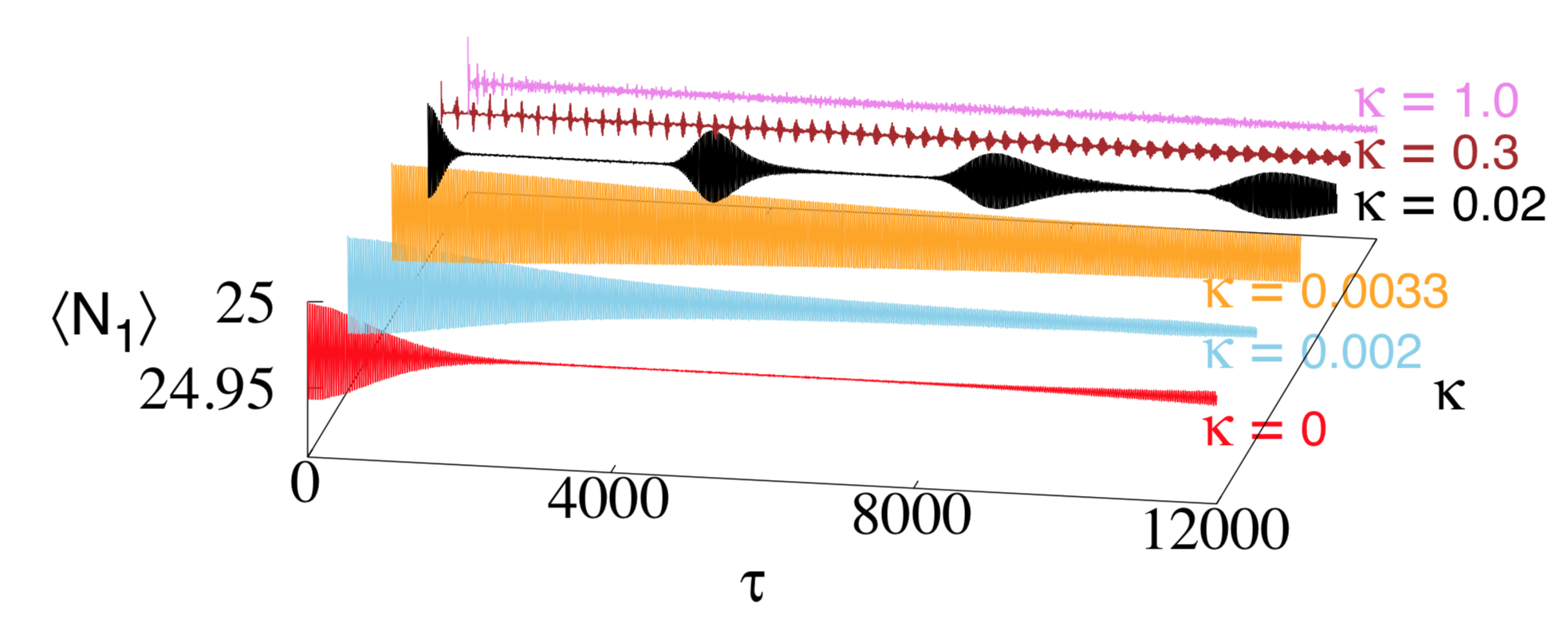}
 \vspace{-0.5ex}
 \caption{$\aver{N_{1}}$ versus time for an initial state $\ket{1;\alpha; \alpha}$,  with ${|\alpha|}^{2} = 25$ and $\chi/\lambda = 5$ for different values of $\kappa$. (Adapted from Fig. 2 of Ref.~\cite{laha4}.)}
 \label{fig:n1_aver}
\end{figure}

For ready reference, it is helpful to start with a quick summary  of the pertinent  features of the system at hand that have been established in earlier work~\cite{laha1,laha4}.  The observable we focus on is the mean photon number of either field.  For definiteness, we consider $\aver{N_{1}}$. The time series of this observable reveals interesting dynamical features. In the strong nonlinearity regime ($\chi/\lambda \gg 1$), $\aver{N_{1}}$ undergoes a  sequence of bifurcations as $\kappa$ is varied from 0 to 1. This is depicted in  Fig. \ref{fig:n1_aver}, which corresponds to the values $|\alpha|^{2} = 25$ and $\chi/\lambda = 5$. When $\kappa = 0$,  $\aver{N_{1}}$ collapses to a constant value  in the time interval $3000 \lesssim \tau \lesssim 9000$, where $\tau$ denotes the dimensionless time $\lambda t$. As $\kappa$ increases to the value  $0.002$, this collapse  is replaced by a `pinched' effect over  the same time interval.  Going on, at $\kappa =  0.0033 \equiv \bar{\kappa}$  there is a significantly larger spread in the range of values of $\aver{N_{1}}$, and the pinch seen for lower values of  $\kappa$ disappears. When $\kappa$ increases beyond  $\bar{\kappa}$, the qualitative behavior of $\aver{N_{1}}$ reverts to its original  form---e.g., the behavior for $\kappa = 0.005$  is very similar to that for $ \kappa = 0.002$.  Within the computational accuracy involved, $\bar{\kappa} = 0.0033$ is identifiable as a special value  of $\kappa$ (for the given values of  $\vert\alpha\vert^{2}$ and $\chi/\lambda$). With an even  further increase in $\kappa$,  the pinched effect returns in the form of a collapse, and an oscillatory pattern in  $\aver{N_{1}}$ sets in. The spacing between successive crests and troughs diminishes with increasing  $\kappa$. This feature persists up to $\kappa = 1$.

For different values of $\kappa$, $\epsilon$-recurrence networks have been constructed~\cite{laha4}  from the corresponding time series of $\aver{N_{1}}$,  setting $|\alpha|^{2} = 25$ and $\chi/\lambda = 5$. In each case, standard network measures such as the average path length, the link density, the clustering coefficients,  the transitivity, the assortativity and the degree distributions have been  examined. It has been shown that  the clustering coefficient and the transitivity  have a maximum value at $\bar{\kappa}$, thus capturing  an important feature of the short-time dynamics. We now proceed to investigate the behavior of the MLEs, recurrence plots and  first-return-time distributions as a function of $\kappa$, with particular reference to the behavior at the special value $\bar{\kappa}$. We also examine  whether the clustering coefficient and transitivity maximize at $\bar{\kappa}$ for other choices of  the network.

\section{\label{sec:time_series}  Time-series  analysis and network properties }
  For each given value of $\kappa$, a long time series  of the mean  photon number $\aver{N_{1}}$ was generated, comprising $N_{\textrm{tot}}$ data points with a scaled time step $\lambda\,\delta t= 1$. From each of these time series, the first 10000 data points were discarded, to obtain a corresponding time series $\{s(i)\}$ where  $1 \leqslant  i  \leqslant   N$ and $N = N_{\textrm{tot}}  -10000$. Hence this set does not include  the short-time dynamics--- in particular,  the bifurcation cascade. The range of values of  $\{s(i)\}$ evidently depends on the specific value of $\kappa$ selected.  Next,  employing the standard machinery of time-series analysis (see, e.g.,~\cite{abarbanel}), a suitable time delay $t_{d}$ was identified,  and an effective phase space of dimensions $d_{\textrm{emb}} \,(\ll N)$ was reconstructed from $\{s(i)\}$.  In this space there are $N' = N - (d_{\textrm{emb}} - 1)t_{d}$ delay vectors 
 ${\bf x}_{j} \,\, (1 \leqslant j \leqslant N')$ given by
\begin{equation}
 {\mathbf x}_{j} = \big[s(j),\, s(j+t_{d}), \cdots,\, 
 s\big(j + (d_{\textrm{emb}}-1)t_{d}\big)\big].
 \label{eqn:delay_vec}
\end{equation}
Of the $d_{\rm emb}$ Lyapunov exponents in the phase space, the maximal Lyapunov exponent (MLE) $\lambda_{\rm max}$ was then  computed using the standard  TISEAN package~\cite{tisean}. 

\begin{figure}
 \centering
 \includegraphics[width=8.5cm, height=3.5cm]{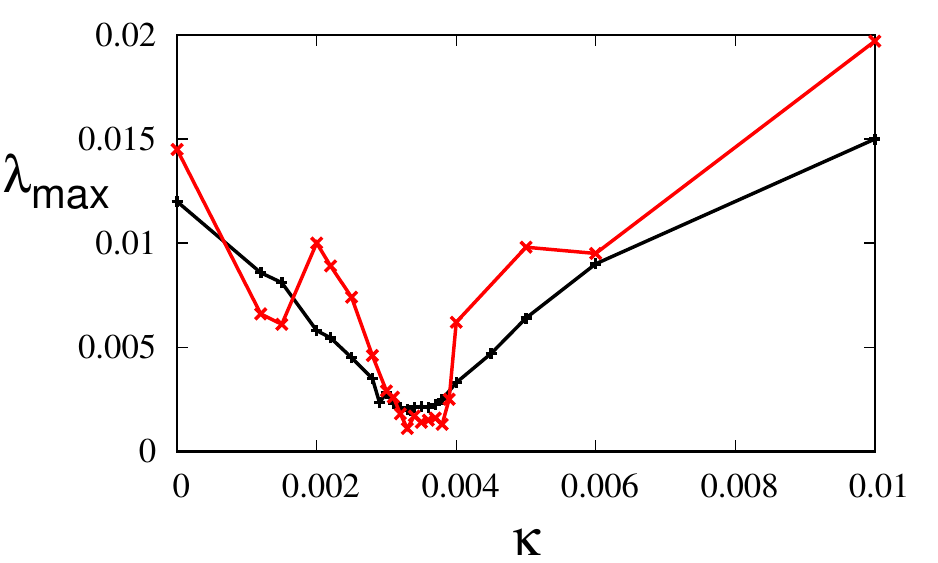}
 \vspace{-0ex}
 \caption{MLE versus $\kappa$ for $N=25000$ (red curve) and $N=3\times 10^{5}$ (black curve). $\vert\alpha\vert^{2} = 25$ and $\chi/\lambda  = 5$. (For interpretation of the colors in the figure(s), the reader is referred
to the web version of this article.)}
 \label{fig:lyap_expo}
\end{figure}

We have compared the manner in which the  MLE  varies with  $\kappa$  for  $N=25000$ and for $N = 3\times10^{5}$ data points (respectively, the red and black curves in Fig. \ref{fig:lyap_expo}). The two  curves are roughly similar, the curve becoming smoother with an increase in $N$. 
\begin{figure}[h]
 \centering
 \includegraphics[width=8.5cm, height=3.5cm]{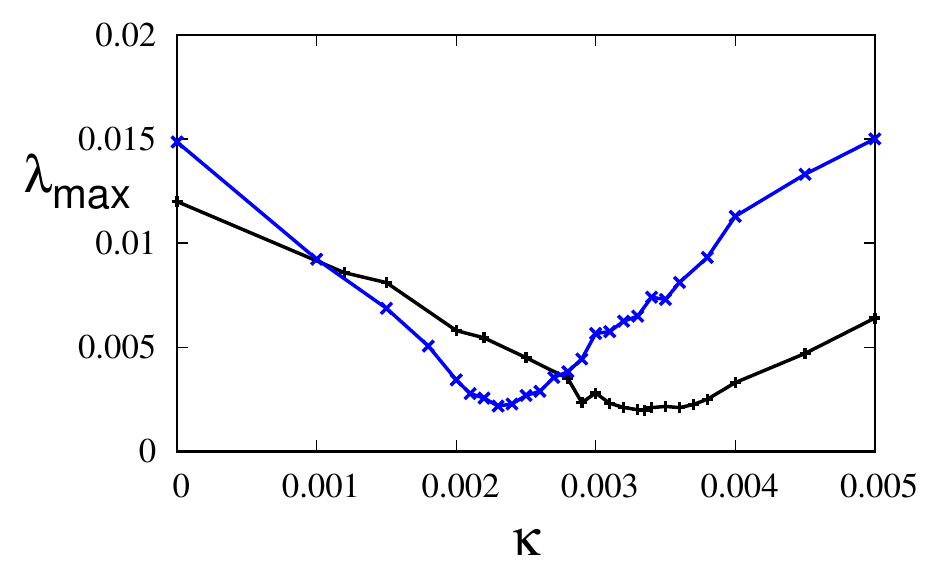}
 \vspace{-0ex}
 \caption{MLE versus $\kappa$ with $N=3\times 10^{5}$, $\chi/\lambda  = 5$. $\vert\alpha\vert^{2} =  25$ (black curve) and $\vert\alpha\vert^{2} = 30$ (blue curve).}
 \label{fig:lyap_expo2}
\end{figure}
\begin{figure*}
\centering
\includegraphics[height=3.50cm,width=4.95cm,angle=-0]{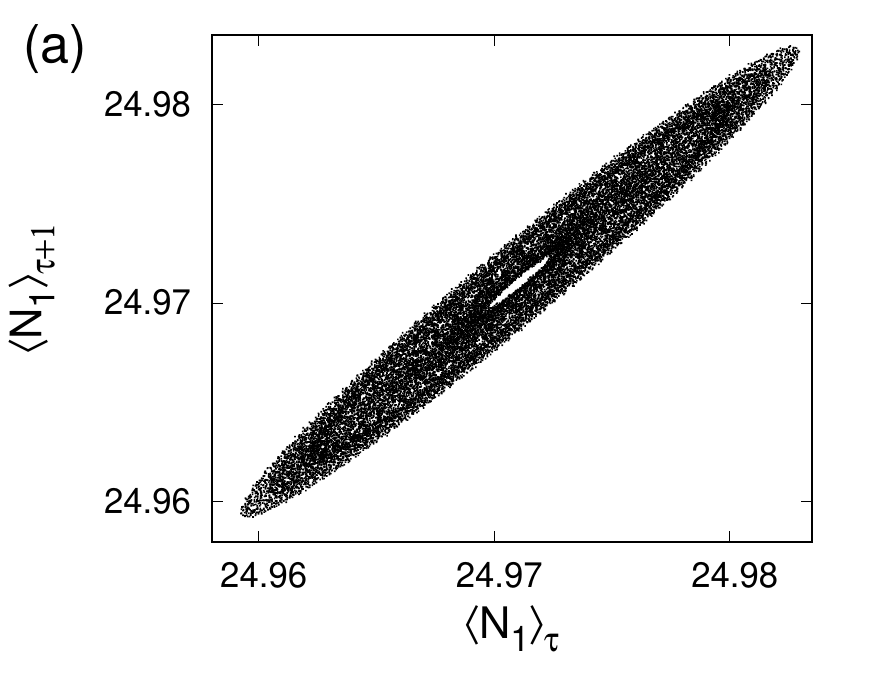}
\includegraphics[height=3.50cm,width=4.95cm,angle=-0]{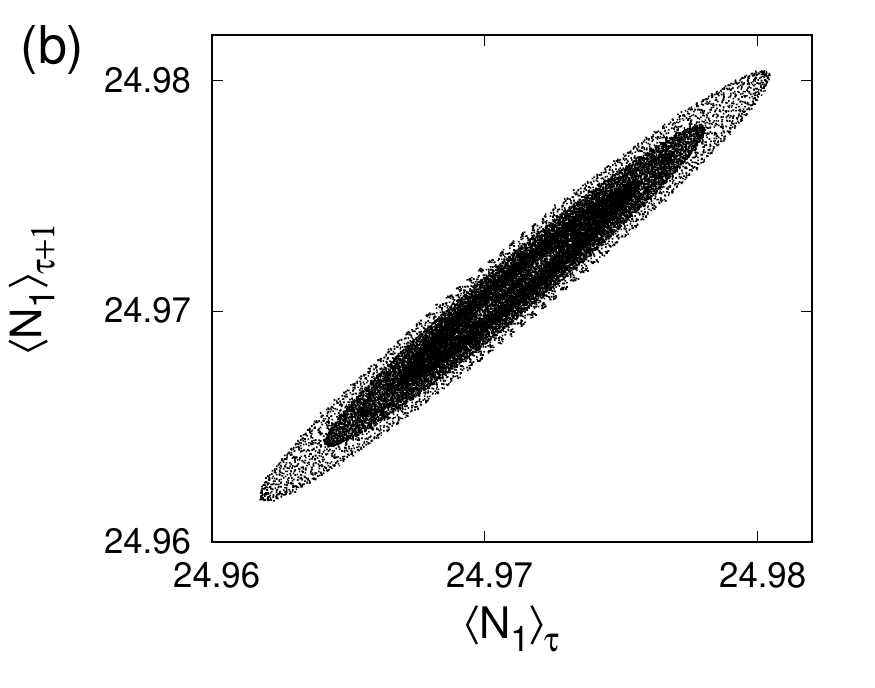}
\includegraphics[height=3.50cm,width=4.95cm,angle=-0]{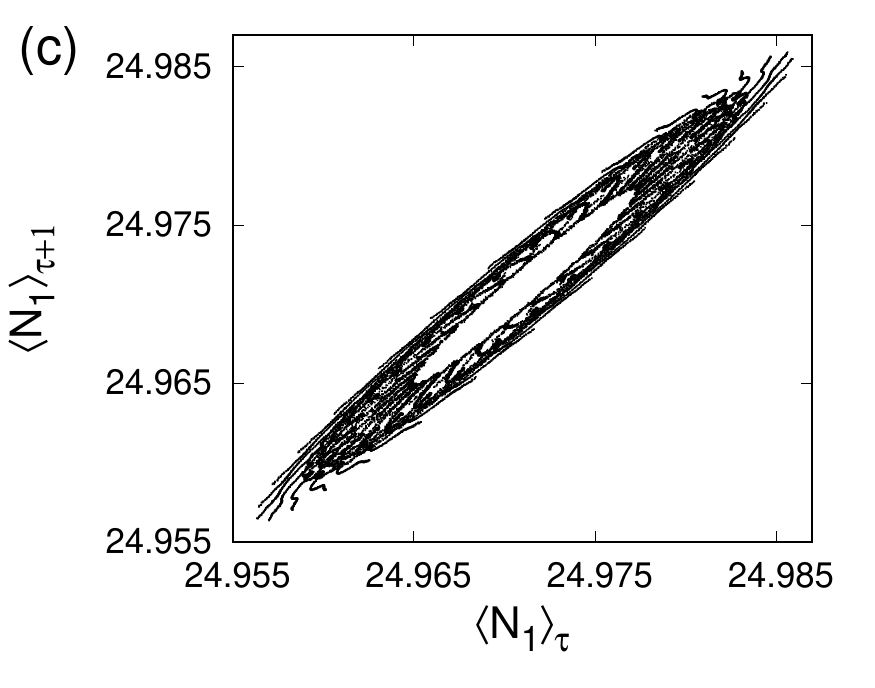}\\
\includegraphics[height=3.50cm,width=4.95cm,angle=-0]{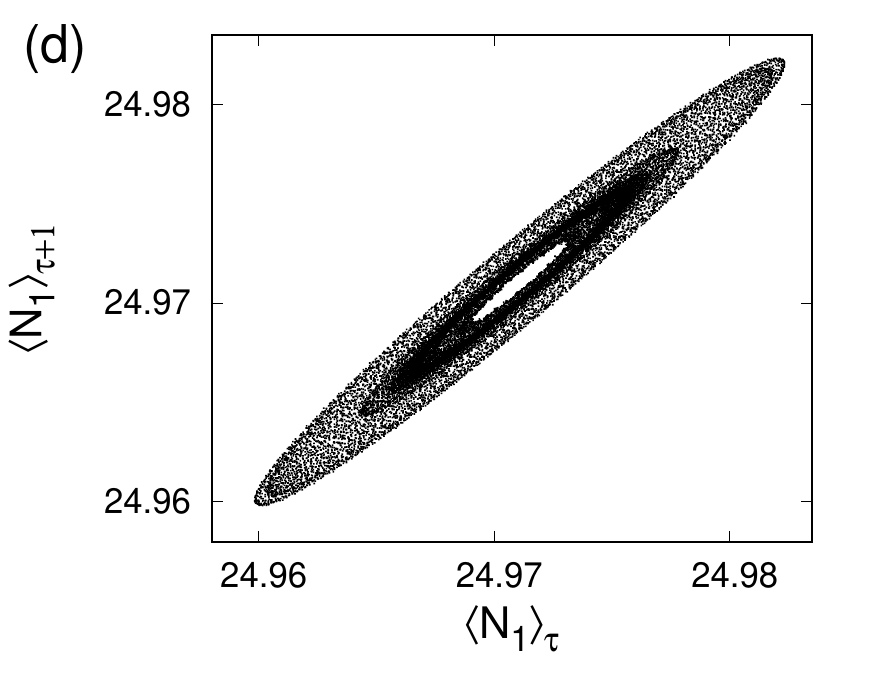}
\includegraphics[height=3.50cm,width=4.95cm,angle=-0]{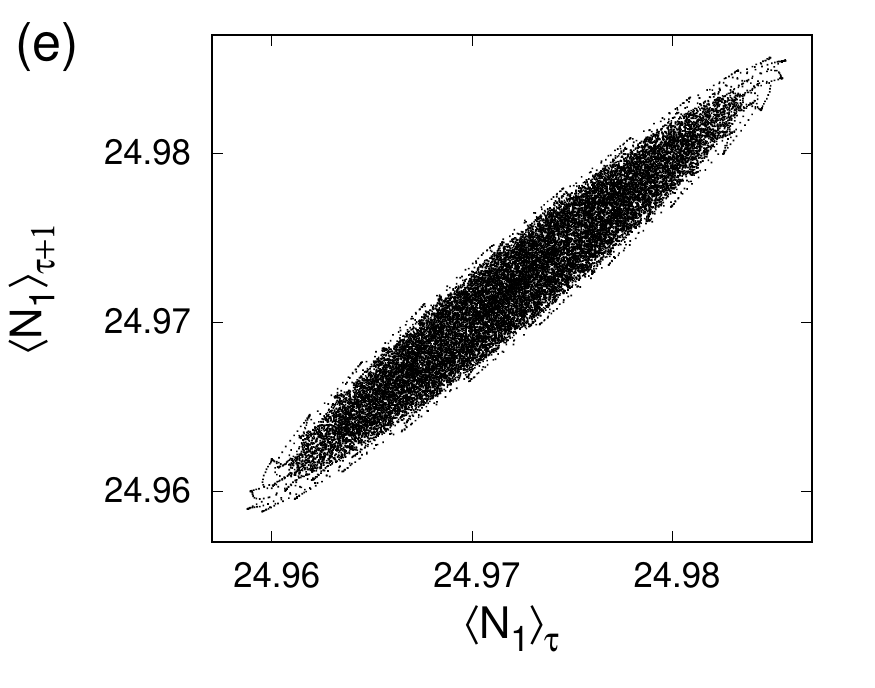}
\includegraphics[height=3.50cm,width=4.95cm,angle=-0]{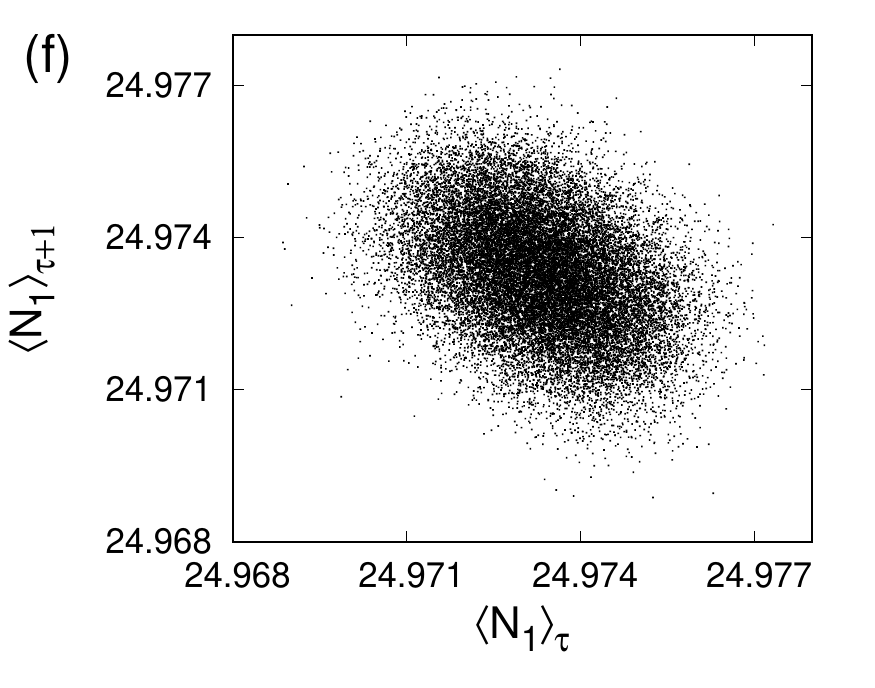}\\
\caption{Return maps of $\aver{N_{1}}$. $|\alpha|^{2} = 25$, $\chi/\lambda = 5$,  and $\kappa = $ (a) 0, (b) 0.002,  (c) 0.0033,  (d) 0.005, (e) 0.02 and (f) 1.}
\label{fig:return_map}
\end{figure*}
\begin{figure*}
 \centering
 \includegraphics[height=3.50cm,width=4.95cm,angle=-0]{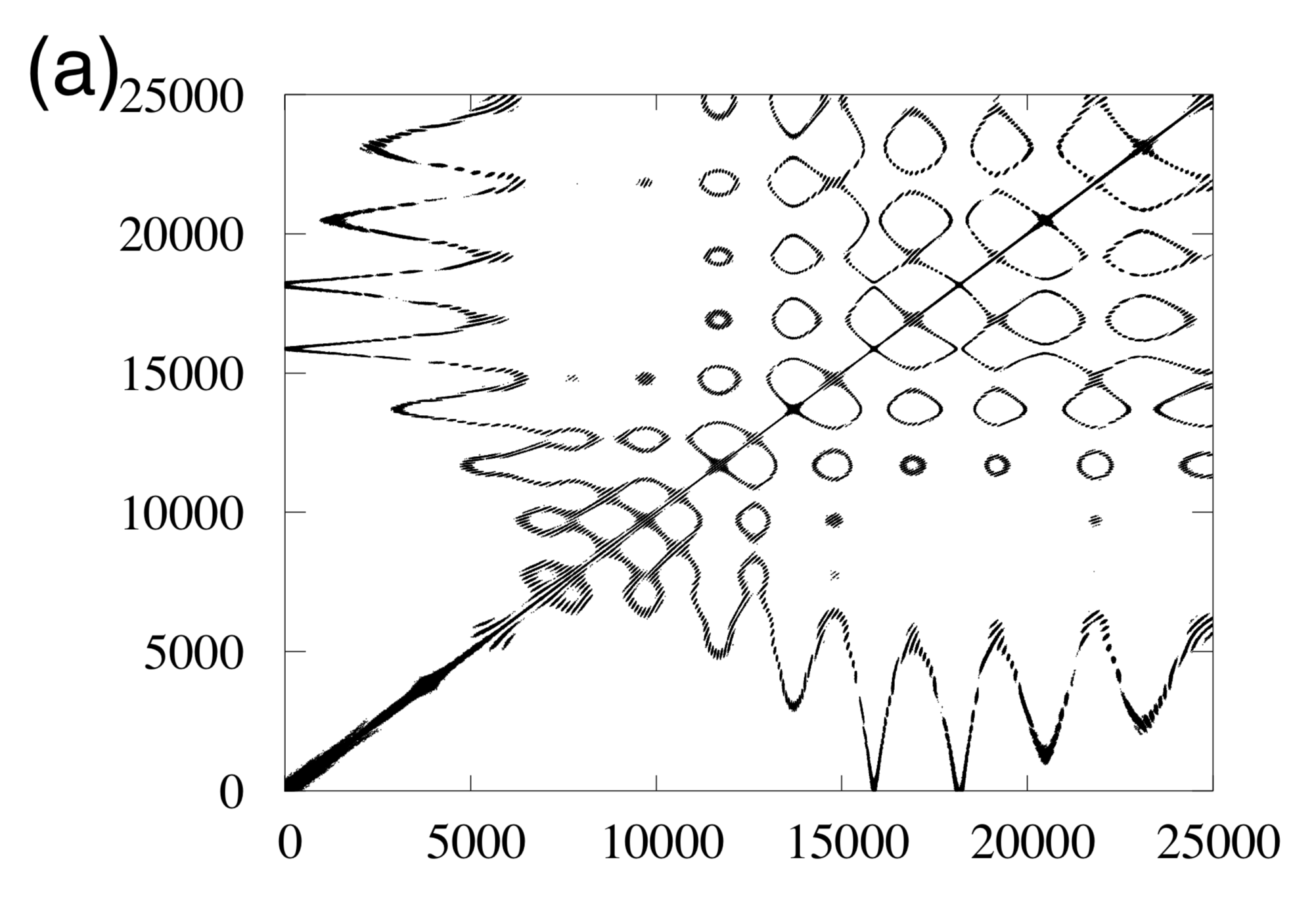}
 \includegraphics[height=3.50cm,width=4.95cm,angle=-0]{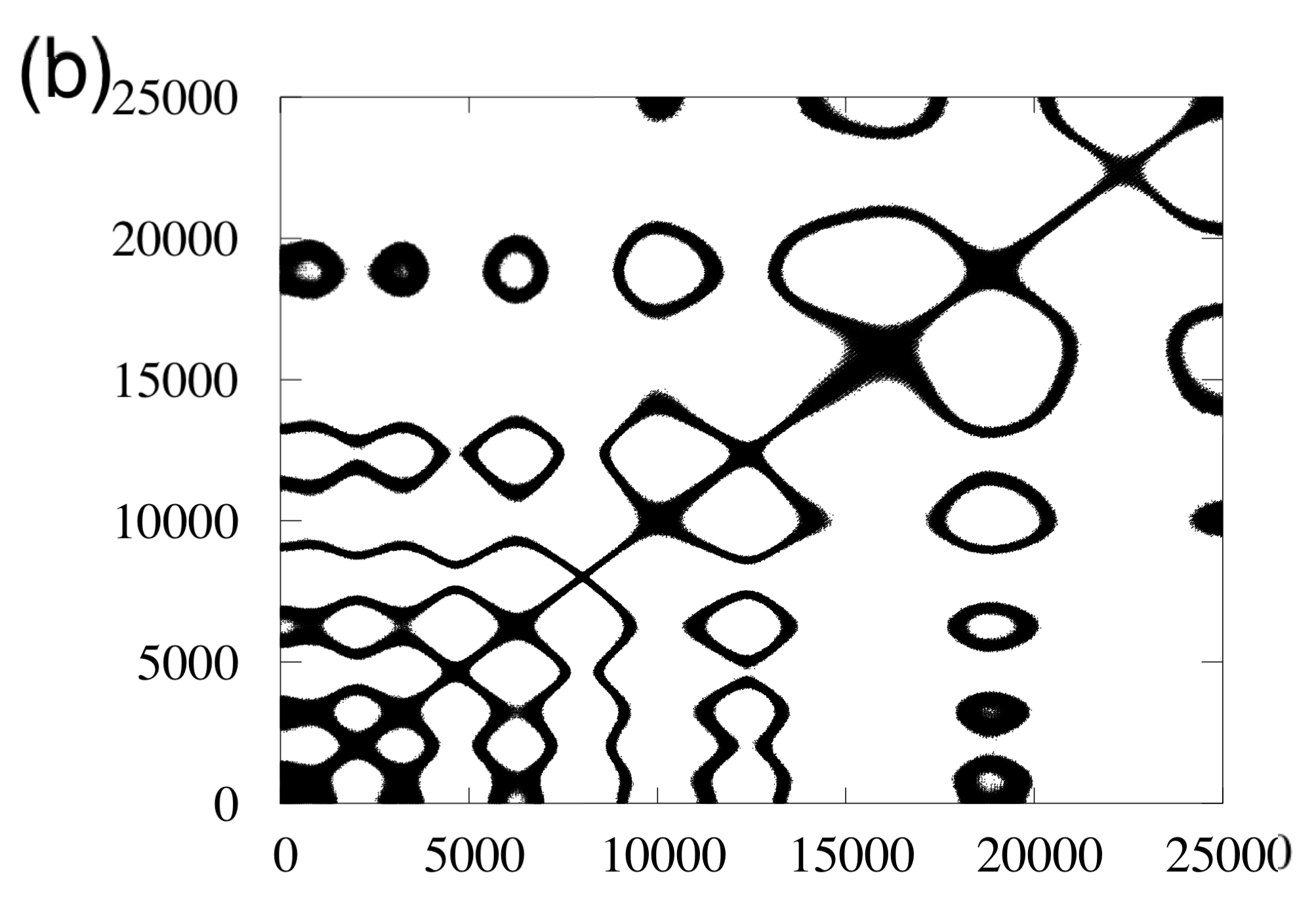}
 \includegraphics[height=3.50cm,width=4.95cm,angle=-0]{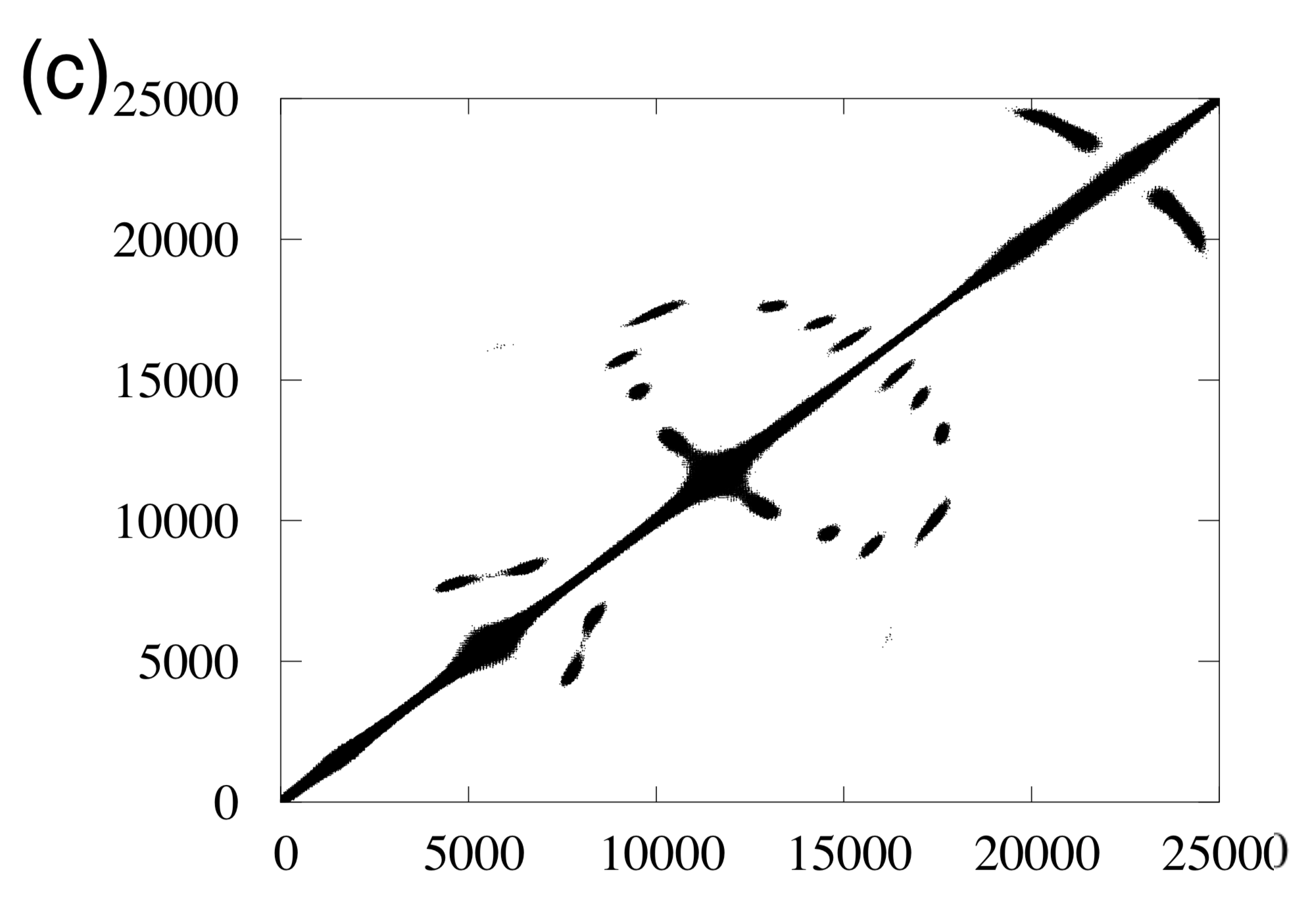}\\
 \includegraphics[height=3.50cm,width=4.95cm,angle=-0]{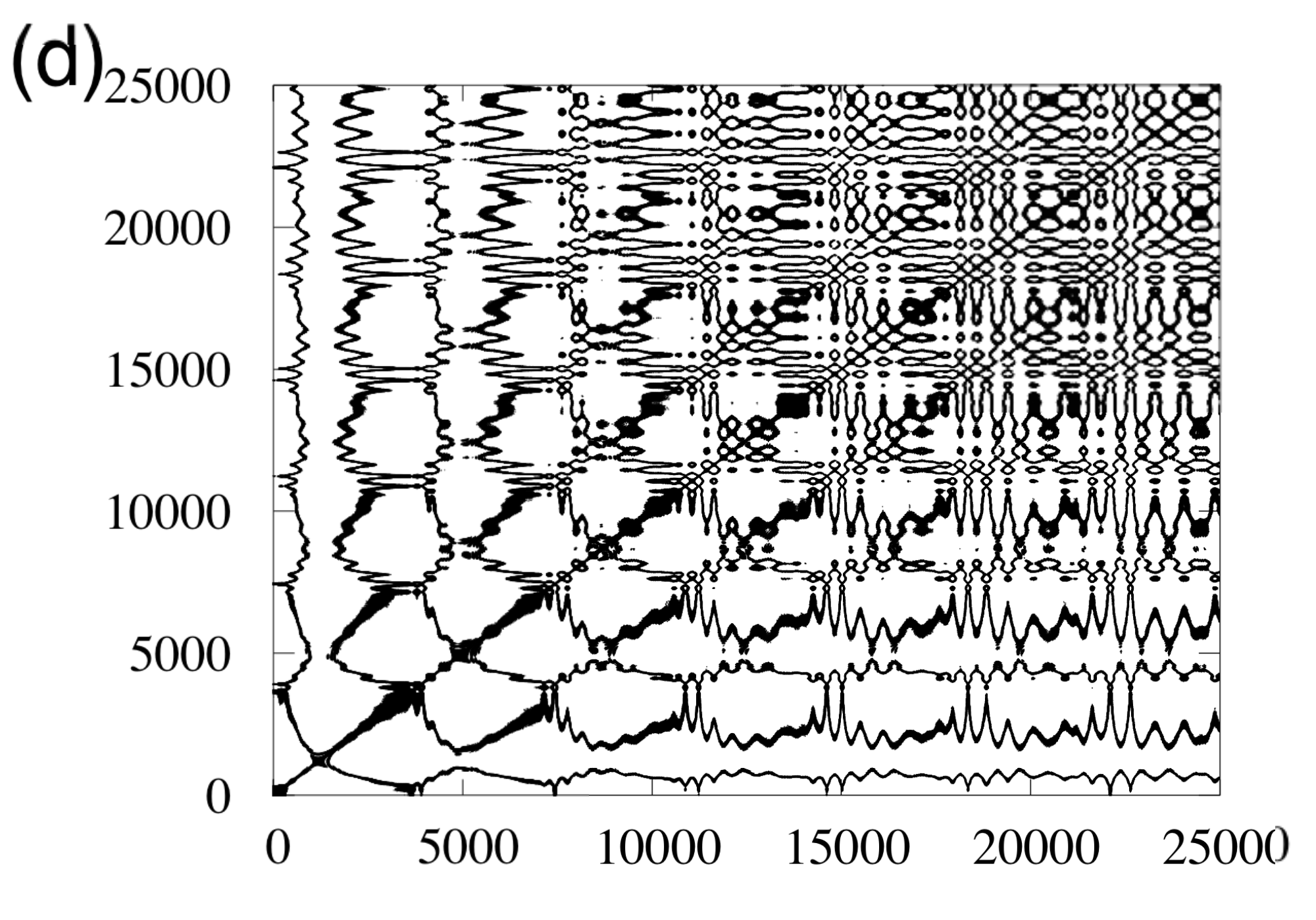}
 \includegraphics[height=3.50cm,width=4.95cm,angle=-0]{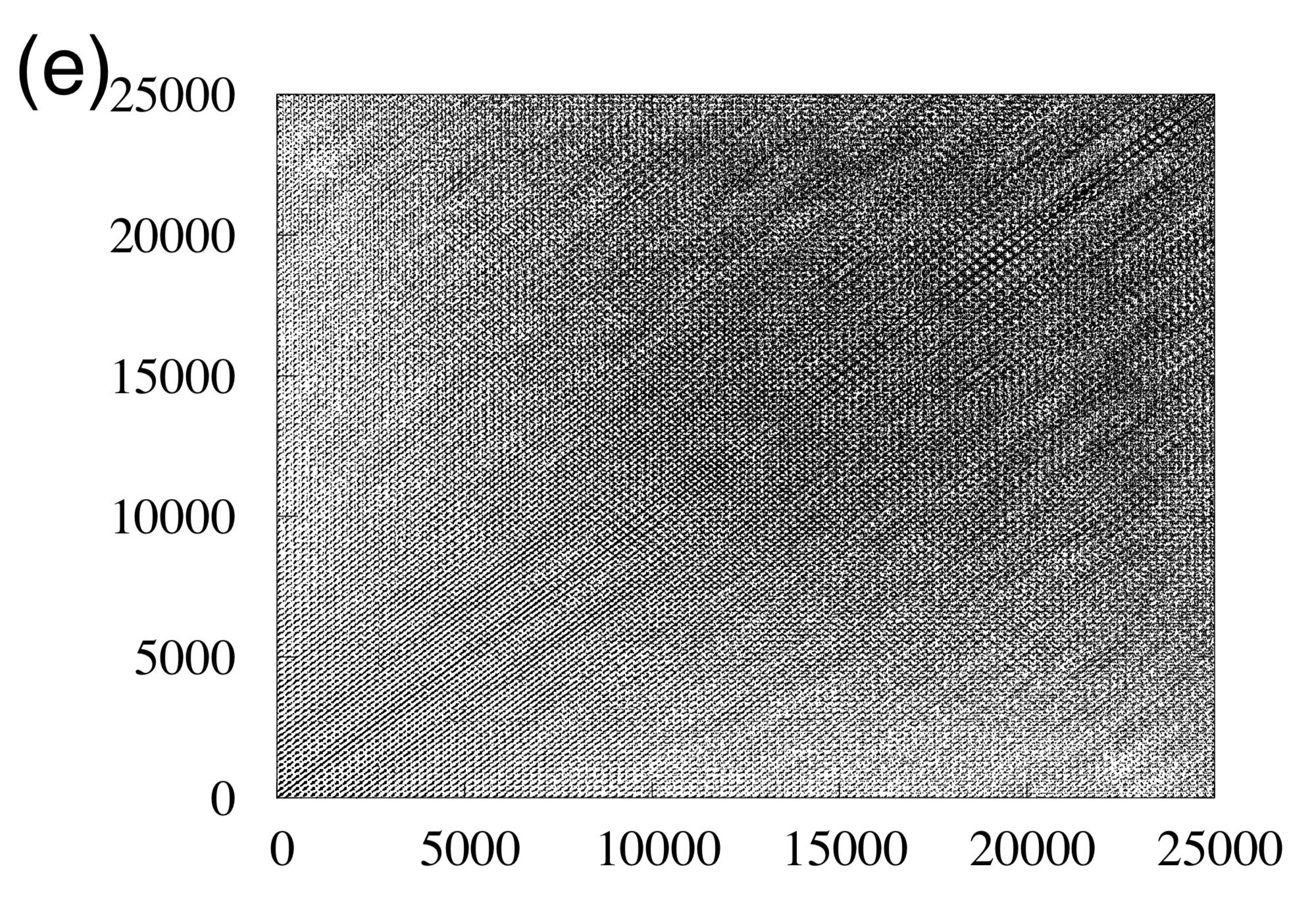}
 \includegraphics[height=3.50cm,width=4.95cm,angle=-0]{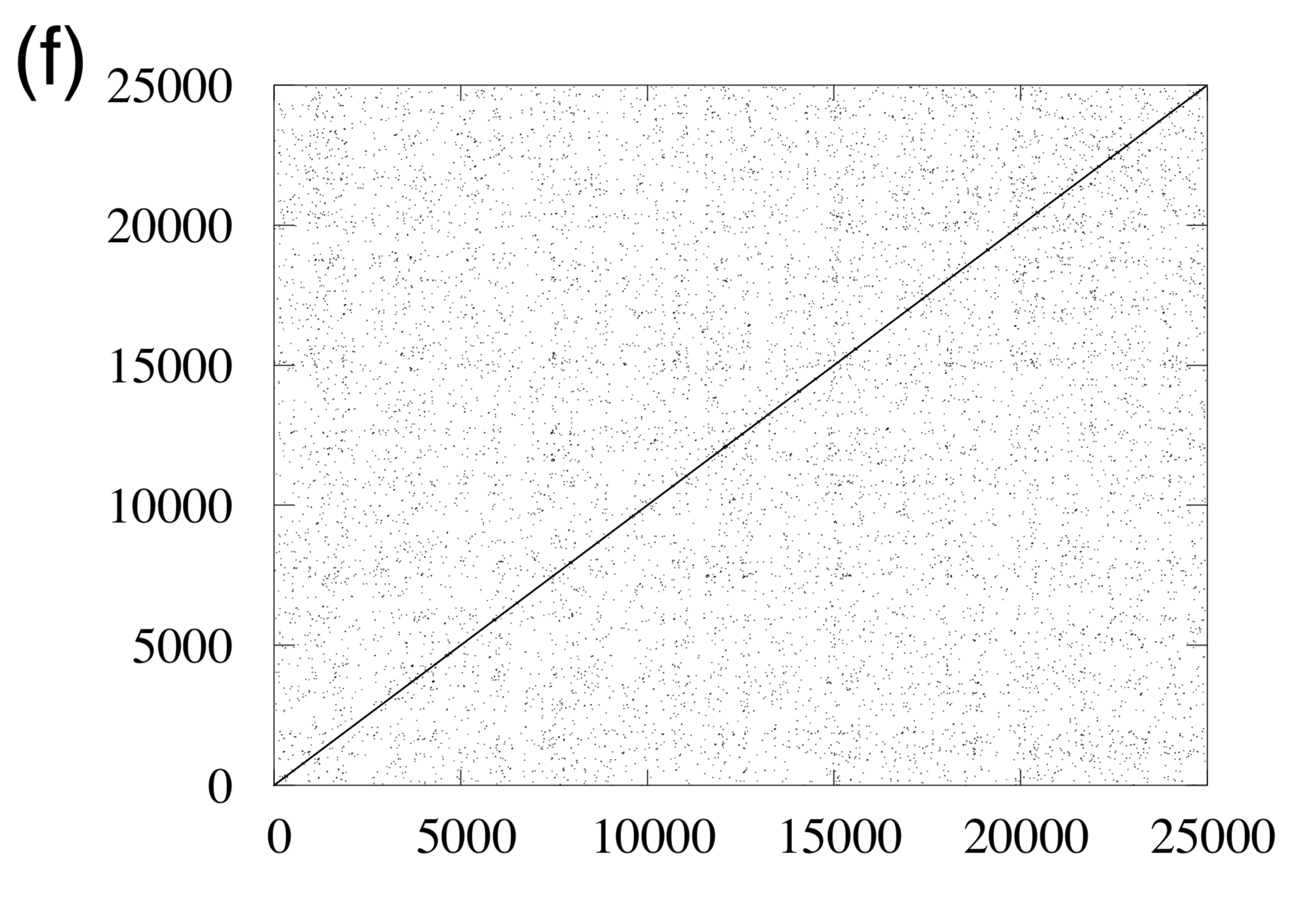}\\
 \caption{Recurrence plots of $\aver{N_{1}}$. $|\alpha|^{2} = 25$, $\chi/\lambda = 5$,  and $\kappa = $ (a) 0, (b)0.002, (c) 0.0033, (d) 0.02, (e) 0.3 and (f) 1.}
 \label{fig:rec_plots}
\end{figure*}
 In particular, we draw attention  to the minimum  value of the MLE in both cases at $\kappa = \bar{\kappa}$. We have  verified that at least 25000 data points are necessary to capture the qualitative behavior of the dynamics in this system. Although the short-time dynamics (including the bifurcation phenomenon around~$\bar{\kappa}$)  has been excluded in these data sets, the clear minimum in  the MLE serves to identify   this  special value of~$\kappa$. We note in passing that the reconstructed dynamics is chaotic,  but only weakly so, as indicated by the small numerical values of~$\lambda_{\rm max}$.

For any given value of $\vert\alpha\vert^{2}$, there is a unique special value $\bar{\kappa}$ with the properties described in the foregoing.  For instance, when $\vert\alpha\vert^{2} = 30$, we find that  $\bar{\kappa} = 0.0024$~\cite{laha4}. As before,  the MLE has a minimum at $\bar{\kappa}$ (the blue curve in Fig. \ref{fig:lyap_expo2}).

We now proceed to examine the return maps, the recurrence plots and the first-return-time distributions for various values of $\kappa$ with  $N = 25000$, $\vert\alpha\vert^{2} = 25$ and $\chi/\lambda  = 5$.  From the return maps of $\aver{N_{1}}$ for different values of $\kappa$  (Figs. \ref{fig:return_map}(a)-(f))  we can see clear signatures of the special value $\bar{\kappa}$.  A  prominent annulus appears  in the return map at 
$\bar{\kappa}$ .  This feature is absent  in the return maps for other values of $\kappa$, although  other substructures and much smaller annuli are present. As $\kappa$ approaches 1, these substructures also disappear,  and the return maps become more space-filling. 
\begin{figure*}
 \centering
 \includegraphics[height=3.50cm,width=4.95cm,angle=-0]{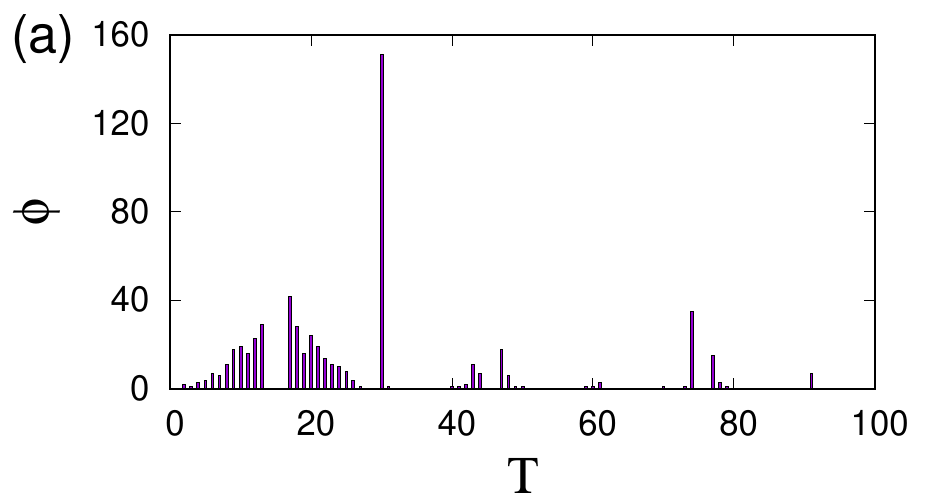}
 \includegraphics[height=3.50cm,width=4.95cm,angle=-0]{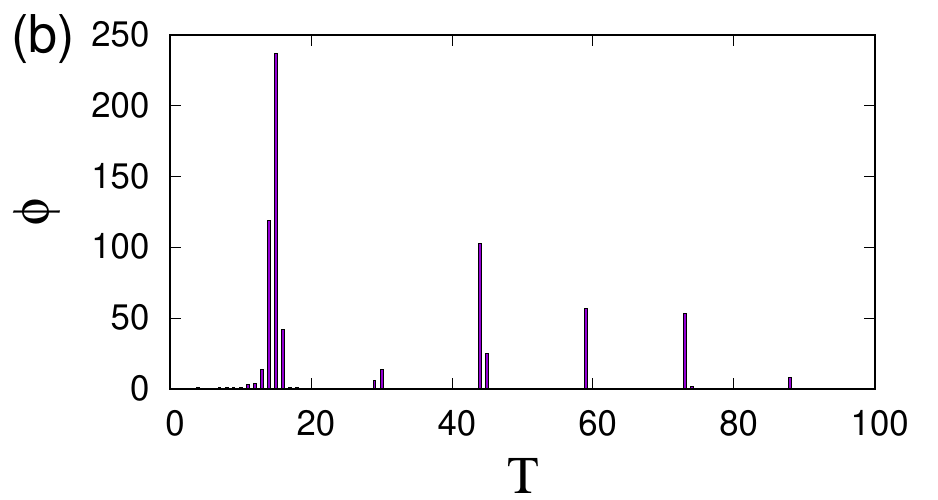}
 \includegraphics[height=3.50cm,width=4.95cm,angle=-0]{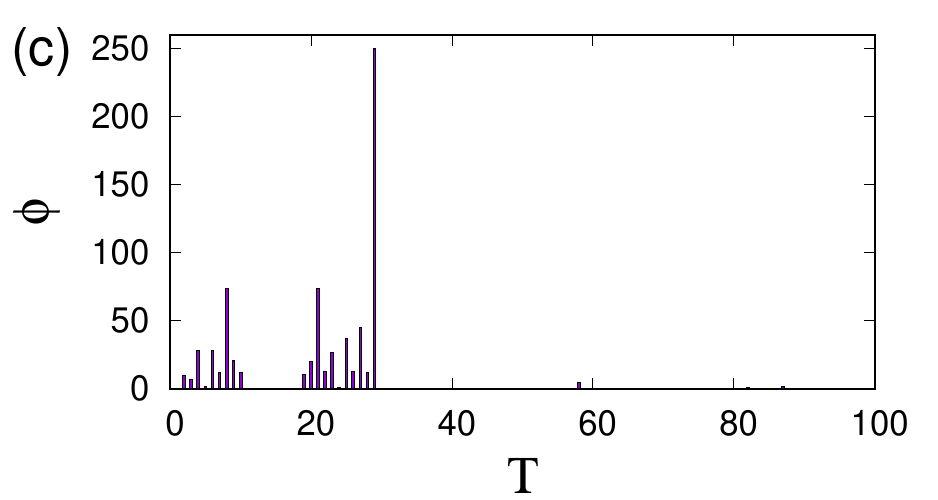}\\
 \includegraphics[height=3.50cm,width=4.95cm,angle=-0]{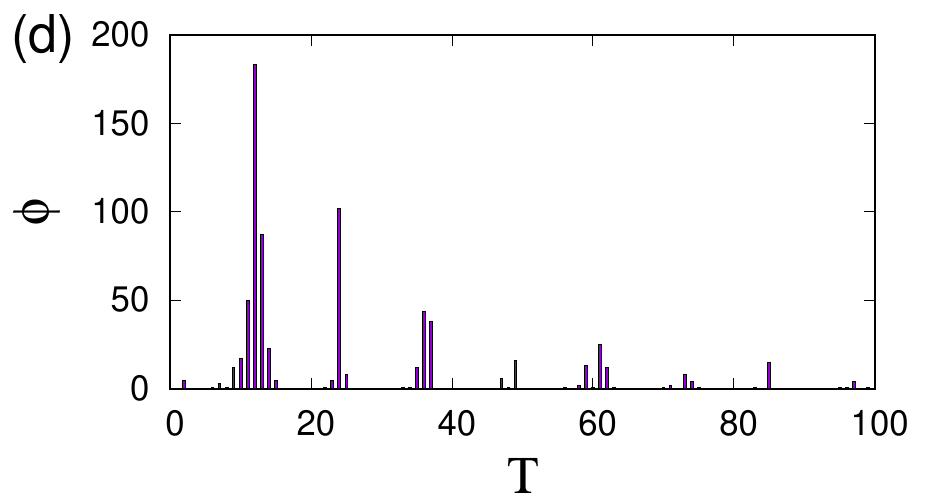}
 \includegraphics[height=3.50cm,width=4.95cm,angle=-0]{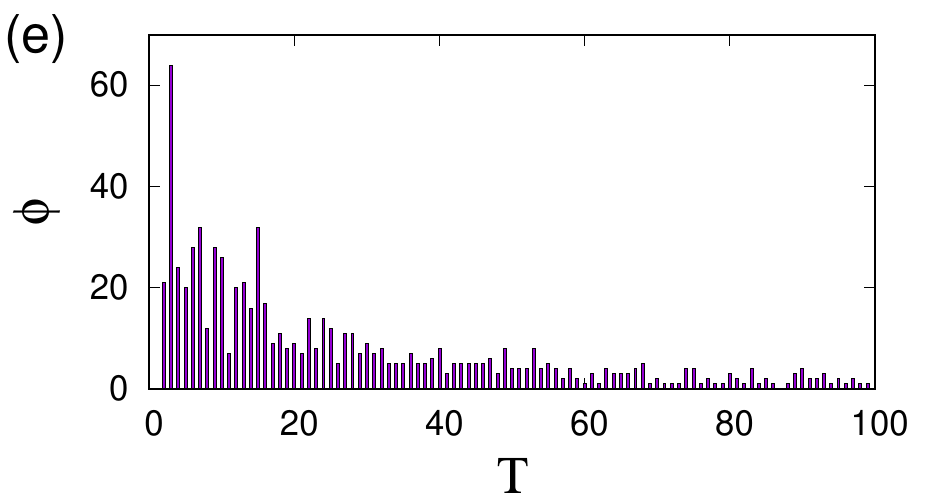}
 \includegraphics[height=3.50cm,width=4.95cm,angle=-0]{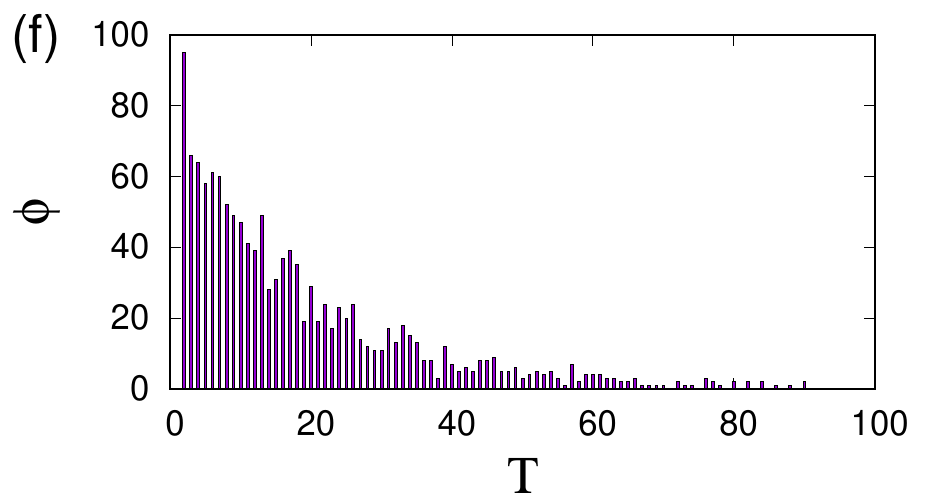}
 \caption{First-return-time distribution of $\aver{N_{1}}$  for 50 equal-sized cells. $|\alpha|^{2} = 25$, $\chi/\lambda = 5$, and $\kappa = $ (a) 0, (b) 0.002, (c) 0.0033, (d) 0.02, (e) 0.3 and (f) 1.}
 \label{fig:first_ret_dist}
\end{figure*}
\begin{figure*}
 \centering
 \includegraphics[height=3.50cm,width=4.95cm,angle=-0]{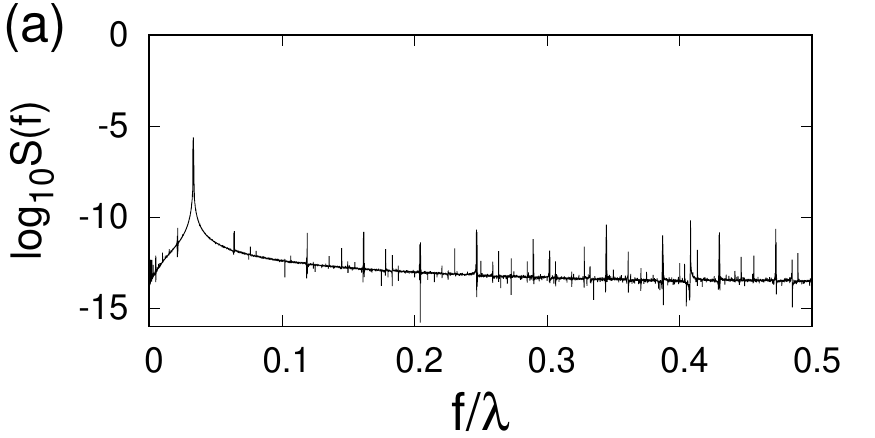}
 \includegraphics[height=3.50cm,width=4.95cm,angle=-0]{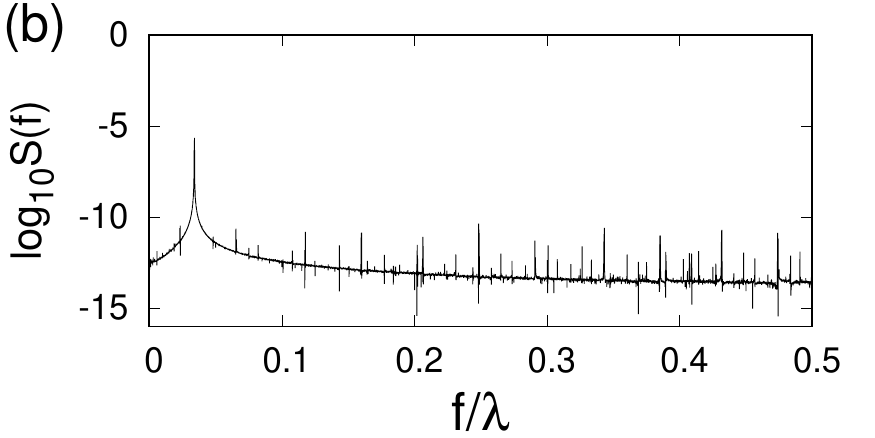}
 \includegraphics[height=3.50cm,width=4.95cm,angle=-0]{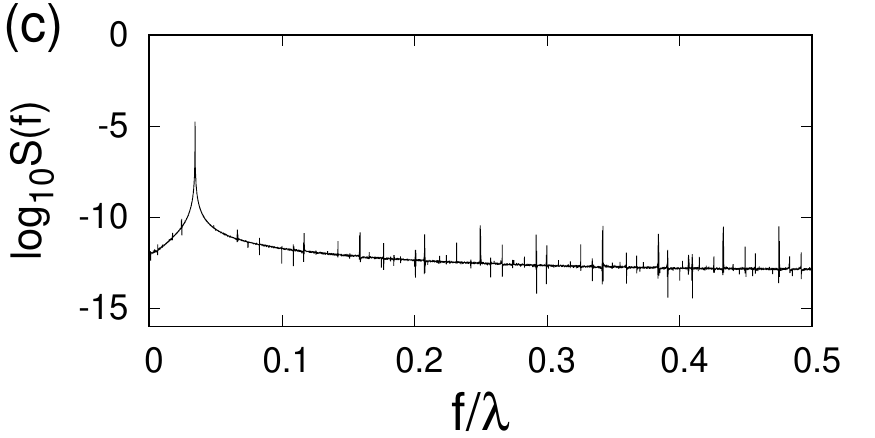}\\
 \includegraphics[height=3.50cm,width=4.95cm,angle=-0]{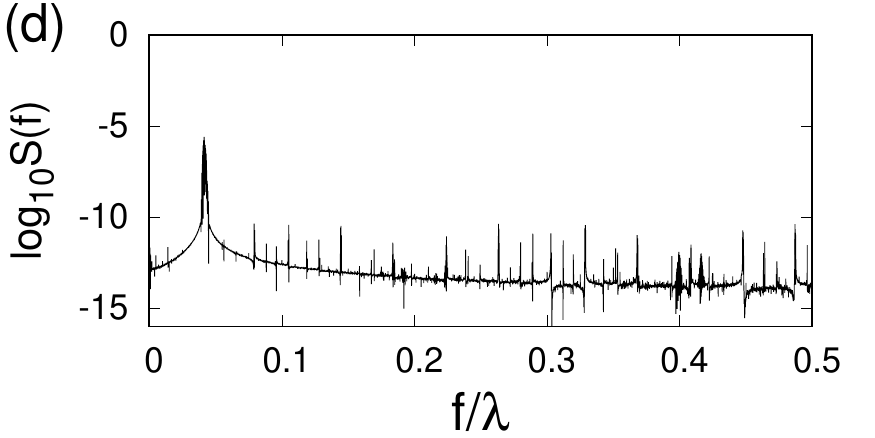}
 \includegraphics[height=3.50cm,width=4.95cm,angle=-0]{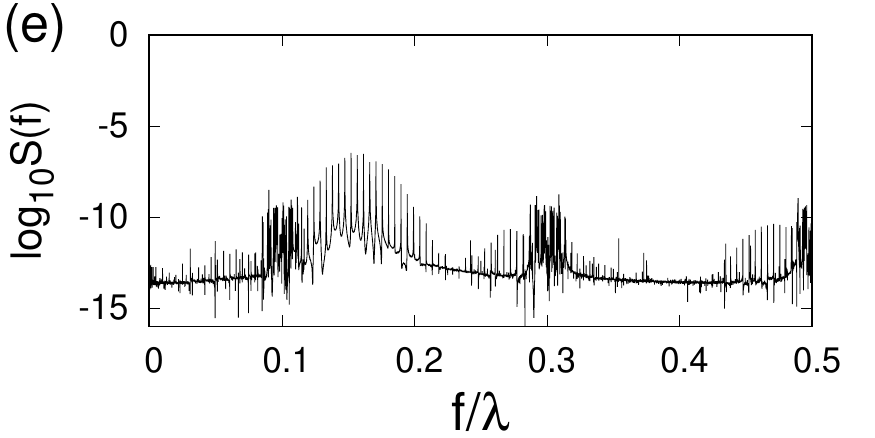}
 \includegraphics[height=3.50cm,width=4.95cm,angle=-0]{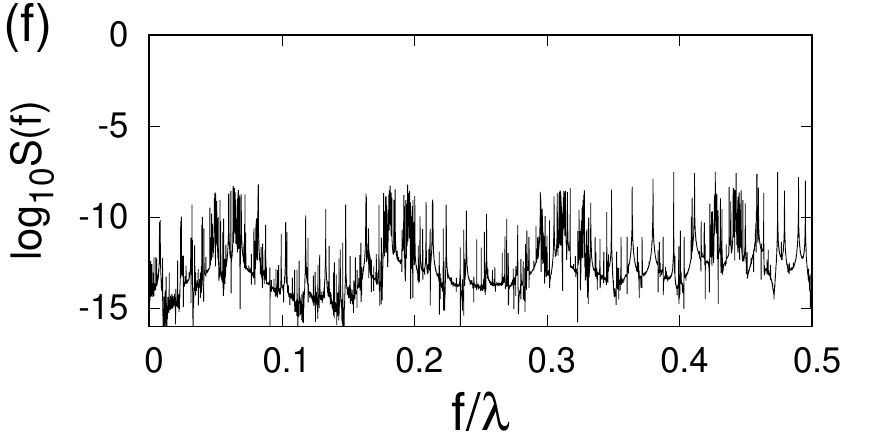}
 \caption{Power spectrum of $\aver{N_{1}}$. $|\alpha|^{2} = 25$, $\chi/\lambda = 5$,  $\kappa = $ (a) 0, (b) 0.002, (c) 0.0033, (d) 0.02, (e) 0.3 and (c) 1.}
 \label{fig:power_spec}
\end{figure*}

Recurrence plots  are standard tools to  understand the dynamics in a reconstructed  phase space \cite{eckmann,thiel1,marwan_rec,marwan_rec_plots}.   Starting from the $(N'\times N')$ recurrence matrix $R$ with elements   
\begin{equation}
 R_{ij} = \Theta(\epsilon -  \parallel \mathbf{x}_{i}-\mathbf{x}_{j} \parallel) 
 \label{eqn:rij}
\end{equation}
(where $\Theta$ denotes the unit step function and $\parallel \cdot\parallel$ is the Euclidean norm), we construct recurrence plots for different values of $\kappa$. In doing so, the threshold parameter~$\epsilon$ must be chosen judiciously. Too small a value of  $\epsilon$ makes the plot very sparse with very few recurrences, while too large a value yields false recurrences. Several  criteria have been proposed in  the literature~\cite{mindlin,zbilut1,zbilut2,thiel2} for choosing an optimal  value of $\epsilon$. We adopt  the criterion proposed recently~\cite{eroglu} in the context of $\epsilon$-recurrence networks. We consider the  $(N'\times N')$ Laplacian matrix $L$ with elements
\begin{equation}
 L = D - R +I, 
 \label{eqn:lij}
\end{equation}
  where $I$ is the unit matrix, $D = {\rm diag} \,(k_{1},  \, \dotsc, \,k_{N'})$ and  $k_{i} = \sum_{j=1}^{N'} R_{ij} - 1$. $L$ is a real symmetric matrix, and each of its row sums vanishes.  It follows from the structure of $L$ that  the eigenvalues  of $L$ are real and non-negative, and  that at least one of them  is zero. Increasing $\epsilon$ upward from zero, we determine the smallest value of $\epsilon$  (denoted by $\epsilon_{c}$)  for which the next eigenvalue of $L$ becomes nonzero. Recurrence plots ($N'\times N'$ grids with elements~$R_{ij}$) are obtained from Eq.\eqref{eqn:rij} with $\epsilon$ set equal to $\epsilon_{c}$, for various values of $\kappa$. These plots highlight  the importance of the special value  $\bar{\kappa}$ (Figs. \ref{fig:rec_plots}(a)-(f)).  It is evident that the plot is distinctly sparse for  $\kappa = \bar{\kappa}$.  For other values of $\kappa$ over the  interval $0 \leqslant \kappa\leqslant 1$, the corresponding recurrence plots are more dense, some with defined structures and some merely  a  set of space-filling points.
  
 We turn, next,  to the recurrence-time statistics, by first coarse-graining the range of values in the time series of  $\aver{N_{1}}$ for different values of $\kappa$  into  equal-sized cells. The distribution of the time of first return  to a generic cell (given that the first time step takes  $\aver{N_{1}}$ to a value outside the initial cell) is obtained.  This  procedure has been carried out for different initial  cells, different cell sizes and for different values of $\kappa$. Typical  first-return-time distributions  are shown in Figs. \ref{fig:first_ret_dist}(a)-(f). For values of $\kappa$ close to $0$, the distributions have a single pronounced spike. As $\kappa$ increases, more spikes appear, and the distributions  gradually tend  to  exponential distributions as $\kappa \rightarrow  1$.  For values of $\kappa$ close to $\bar{\kappa}$ on either side of that value, all the other spikes in the distribution lie to the right of the most pronounced spike. At $\kappa = \bar{\kappa}$, however, all these spikes occur to the left of the most pronounced one. The recurrence-time distribution corresponding to the value $\bar{\kappa}$ is thus distinctly different, qualitatively, from that for all other values of $\kappa$. 
 
 We have verified numerically that  the power spectrum is not a good indicator  of the special value $\bar{\kappa}$. For completeness however, the corresponding  power spectra  are shown in Figs. \ref{fig:power_spec}(a)-(f)). 

In \cite{laha4} we constructed an $\epsilon$-recurrence network from the time series under discussion. In such a network, each vector in the reconstructed phase space is treated as a node of a network, and  two nodes are connected if the distance between them is less than  the value $\epsilon_{c}$ defined in the foregoing.  Among other measures we considered the link density (LD), the clustering coefficient (CC) and the transitivity ($\mathcal{T}$)~\cite{kurths_phys_rep,boccaletti, strogatz} of the network. For ready reference we quote the definitions of these quantities. The link density LD of a network of $N'$ nodes is defined as 
\begin{equation}
 \text{LD} = [N'(N'-1)]^{-1} \sum_{i}^{N'} k_{i},
 \label{eqn:apl}
\end{equation}
where $k_{i}= \sum_{j=1}^{N'} A_{ij}$ is now the degree of the  node $i$ in terms of the elements  $A_{ij} = R_{ij} - \delta_{ij}$ of the adjacency matrix $A$. 
The local clustering coefficient, which measures the probability that two randomly chosen neighbours of a given node $i$ are directly connected, is defined as
\begin{equation}
 C_{i} = [k_{i}(k_{i}-1)]^{-1}  \sum_{j,k}^{N'} A_{jk} \, A_{ij}\, A_{ik}.
 \label{eqn:lcc}
\end{equation}
The global clustering coefficient CC is the arithmetic mean of the local clustering coefficients taken over all the nodes of the network.
The transitivity $\mathcal{T}$ of the network is defined as 
\begin{equation}
 \mathcal{T} =   \frac{\sum_{i,j,k}^{N'}   A_{ij}\,A_{jk}\, A_{ki}}{\sum_{i,j,k}^{N'} A_{ij} \, A_{ki}}.
 \label{eqn:transitivity}
\end{equation}
For the effective time series with  $N= 25000$, it was shown that  the CC and $\mathcal{T}$ attain a maximum at $\kappa = \bar{\kappa}$ (see, e.g., Fig.~6 of Ref.~\cite{laha4}). We now establish that, even with a very different and considerably simpler procedure for choosing $\epsilon_{c}$, the CC and $\mathcal{T}$ enable us to single out  the special value $\bar{\kappa}$: the threshold $\epsilon_{c}$  is just the  value of $\epsilon$ for which the link density has an optimal value.
\begin{figure}[h]
 \centering
 \includegraphics[height=3.5cm, width=8cm]{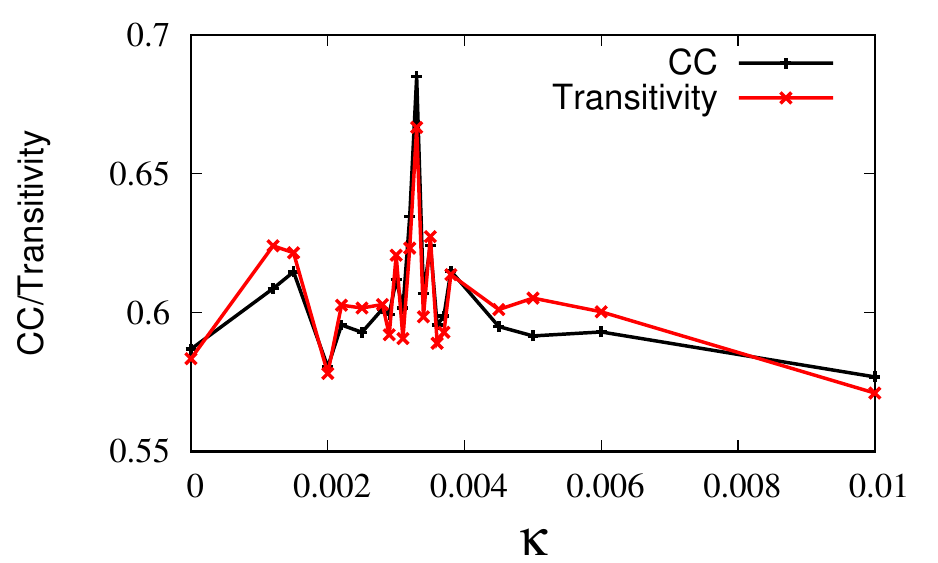}
 \caption{$\epsilon$-recurrence network: clustering coefficient and transitivity versus  $\kappa$ for $\vert\alpha\vert^{2} = 25$ and $\chi/\lambda = 5$.}
 \label{fig:tr_cc_ld}
\end{figure}
  (It has been suggested~\cite{kurths_phys_rep} that this value of the LD  should be  below 0.05.)  We  find that in this procedure of choosing an $\epsilon_{c}$, too, both the CC and $\mathcal{T}$  are maximum at $\kappa = \bar{\kappa}$ (Fig. \ref{fig:tr_cc_ld}). This conclusion is supported by extensive numerical results for a range of values of $\vert\alpha\vert^{2}$ and  LD.  We conclude that an $\epsilon$-recurrence network and the straightforward calculation of its CC and $\mathcal{T}$ provide the simplest means of identifying the special value of the bifurcation parameter $\kappa$ in the model under study.

\section{\label{sec:conclusion} Concluding remarks}
We have investigated, using the machinery of nonlinear time-series analysis, the dynamics of a tripartite quantum system comprising a $\Lambda$-atom interacting with two initially coherent radiation fields with inherent nonlinearities and an intensity-dependent field-atom coupling (IDC). In  earlier work, the short-time dynamics of the mean photon number $\aver{N_{1}}$ was shown to display a sequence of bifurcations as a function of the IDC parameter $\kappa$, in which $\aver{N_{1}}$ shifts from  collapse to a constant value to oscillatory behavior. A special value $\bar{\kappa}$ of the parameter separates the two kinds of behavior. An  $\epsilon$-recurrence network constructed from the time series showed that  the clustering coefficient and the transitivity also displayed signatures of this special value  $\bar{\kappa}$. 

 In this Letter, we have  examined several other relevant aspects of the time series, including the maximum Lyapunov exponents, the return maps, the recurrence plots and the first-return-time distributions.  We have established that the existence of the special value  $\bar{\kappa}$ is mirrored in each of these quantities, and the corresponding distinguishing feature has been identified in each case.  We have further shown that,  independent of the details of the network considered, the clustering coefficient and transitivity carry distinct and easily identifiable signatures of~$\bar{\kappa}$. 
 
 The novelty and importance of this investigation lie in the interesting link established between short-time bifurcation phenomena, on the one hand, and on the other,  network parameters, indicators from dynamical systems  theory such as the qualitative features of recurrences, and quantifiers such as the maximal Lyapunov exponent. The model  we have studied  is a generic one, and the results suggest that the inferences drawn in this case  could be generalized to a wide class of quantum systems. More detailed investigations in this regard would be a useful direction of future research, particularly for understanding possible links between group-theoretic structures on the one hand and the dynamics of quantum observables over short and long time scales on the other.
 
 Nonlinear quantum optics experiments, i.e., attempts  to realize photon-photon interactions through {\it nonlinear coupling of photons to an atom}  have gained tremendous importance  in recent years (see, e.g.,~\cite{chang}).   
 The primary motivation is the belief that, in this regime of quantum optics,  the performance of devices made for quantum technologies and quantum information processing  would be far superior to that of  devices based on classical optics, linear quantum optics, and even nonlinear optics with weak  atom-photon coupling.   
 The main challenge lies in achieving strong nonlinear interactions even with low intensity light. The fact that this target can be achieved through specific kinds of atom-photon coupling is what validates the model we have used, from the point of view of potential experiments. The coupling should clearly be tunable to take into account  
 different levels of light intensity in experiments: in our 
 model, the parameter $\kappa$ multiplies the photon number operator (or intensity). Further, it is known that  
 a  model comprising a {\em two}-level atom interacting with  a photon through appropriate couplings has  limitations, in the sense that  the electronic states of the atom have short lifetimes which prevents them from retaining `memory' of their interactions with the photon over longer  time intervals. This makes it necessary for  at least two photons to arrive simultaneously at the atom, but that leads to spatial entanglement between the two photons which complicates the  results. Hence, a three-level 
  or multi-level atom with two radiation fields which mediate two different transitions is recognized by experimenters as the set-up required. The simplest of course would be the `workhorse', namely the $\Lambda$ atom interacting with two fields, which is the model we have employed. 
 
In this Letter we have explored the   complexity of quantum dynamics in a model which is a {\em paradigm}  of the general features and possibilities  we seek to investigate and categorize.   Now, a detailed body of knowledge has been built up regarding complex behavior in 
{\em classical}  dynamical systems with the help of  representative models, in particular,   low-dimensional maps and flows. The latter  are caricatures of real physical systems, and are not in general faithful representations of actual experimental situations. Nevertheless, their detailed analysis yields very valuable information on the essential and salient features of the dynamics they are intended to demonstrate. A similar objective validates 
the kind of theoretical analysis we have reported  
here on paradigmatic quantum mechanical systems.

\appendix

\section{\label{appendix_state_vector}The state vector}
The Hamiltonian defined in Eq.\eqref{eqn:lambda_hamiltonian} can be written as the sum $H_{0} + H_{1}$, where ${H}_{0}  =  \sum_{i=1}^{2}  \Omega_{i} N_{i}^{\rm tot}  +  \omega_{3} I$ (merely introduces a phase factor in the time evolved state), and
\begin{align}
 {H}_{1}  =   \sum\limits_{i=1}^{2} \Big[\chi a_{i}^{\dagger 2} a_{i}^{2} - \Delta_{i} \sigma_{ii} 
                 + \lambda \{ a_{i}f(N_{i}) \sigma_{3i}  +  f(N_{i})a^{\dagger}_{i} \sigma_{i3} \}\Big].
\label{eqn:ham_h1}
\end{align}
Here $I  =  \sum_{j=1}^{3} \sigma_{jj}$, $N_{i}^{\rm tot}  =  a_{i}^{\dagger} a_{i}  -  \sigma_{ii} \;(i=1,2)$  are constants of the motion,  and  the two detuning parameters $\Delta_{i} = \omega_{3} - \omega_{i} - \Omega_{i}$. For convenience, we set $\Delta_{i} = 0$.
 
 The coefficients $A_{nm}(t)$, $B_{nm}(t)$, and $C_{nm}(t)$ in Eq.\eqref{eqn:two_mode_lambda_interaction_state} satisfy the following coupled equations 
\begin{align}
 \dot{A}_{nm}(t)  &=  (V_{11} + V_{22}) A_{nm}(t) + f_{1} C_{nm}(t), \\[2pt]
 \dot{B}_{nm}(t)  &=  (V_{12} + V_{21}) B_{nm}(t) + f_{2} C_{nm}(t), \\[2pt]
 \dot{C}_{nm}(t)  &=  (V_{12} + V_{21}) C_{nm}(t) + f_{1}A_{nm}(t) + f_{2}B_{nm}(t),
\label{eqn:dot_a_b_c}
\end{align}
with 
\begin{align}
 V_{11}  &=  \chi\, n (n-1), \q V_{12}  =  \chi (n-1)(n-2), \\
 V_{21}  &=  \chi\, m(m+1), \,\,\, V_{22}  =  \chi m (m-1), \\
 f_{1}     &=  \lambda\, n^{1/2} f(n), \,\,\,\,\,\,\,\,\, f_{2}  =  \lambda (m+1)^{1/2} f(m+1).
\label{eqn:two_mode_lambda_f1f2}
\end{align}
Since the atom is initially in $\ket{1}$, it cannot make a transition to $\ket{3}$ if $n = 0$. Hence $A_{0m}(t)  = 1, \;B_{0m}(t)  = C_{0m}(t)  = 0$ for all $m$.  When  $n, m \geq 1$,  we find  
\begin{subequations}
\begin{align}
 A_{nm}(t)  &=  \frac{1} {f_{1}f_{2}} \sum\limits_{j=1}^{3}  \Big[ (\mu_{j}  +  V_{12}  +  V_{22})(\mu_{j}  +  V_{12}  +  V_{21})  \nonumber \\
                  &\hspace{3cm}   -  f_{2}^{2} \Big] b_{j}\, e^{i \mu_{j} t}, \\
 B_{nm}(t)  &=  \sum\limits_{j=1}^{3} b_{j} e^{i \mu_{j} t},\\
C_{nm}(t)  &= -  \frac{1} {f_{2}} \sum\limits_{j=1}^{3}  (\mu_{j}  +  V_{12}  +  V_{21}) b_{j}\, e^{i \mu_{j} t}.
\label{eqn:two_mode_lambda_c}
\end{align}
\end{subequations}
Here, $f_{1}$ and $f_{2}$ capture the effect of $f(N)$ in the Hamiltonian on the state of the system.  Further, for $j  =  1, 2, 3$,
\begin{equation}
 \mu_{j}  =  - {\textstyle\frac{1}{3}}\,x_{1}  +  {\textstyle\frac{2}{3}} \,(x_{1}^{2} -  3 x_{2})^{1/2} \cos \,\big\{\theta  +  {\textstyle\frac{2}{3}} (j-1) \pi  \big\}, 
\label{eqn:mu_j}
\end{equation}
and 
\begin{equation}
 \theta  =  {\textstyle\frac{1}{3}}  \cos^{-1} \Big\{ [9 x_{1} x_{2} -  2 x_{1}^{3}  -  27 x_{3}]
\big/[2 ( x_{1}^{2}  -  3 x_{2} )^{3/2}]\Big\},
\label{eqn:theta}
\end{equation}
with
\begin{align}
 x_{1} &=  V_{11}  +  2 V_{12}  +  V_{21}  +  2 V_{22} ,\\
 x_{2} &= (V_{12}  +  V_{21}) (V_{11}  +  V_{12}  +  2 V_{22})  \nonumber \\
          &\qq+ (V_{12}  +  V_{22}) (V_{11}  +  V_{22})  - f_{1}^{2}  -  f_{2}^{2} ,\\
 x_{3} &= (V_{12}  +  V_{21}) \left[ (V_{12}  +  V_{22}) (V_{11}  +  V_{22})  -  f_{1}^{2} \right]  \nonumber \\
          &\qq -  f_{2}^{2} (V_{11}  + V_{22}), 
\label{eqn:two_mode_lambda_x3}
\end{align}
and
\begin{equation}
  b_{j}  = f_{1} f_{2}/[(\mu_{j}-\mu_{k}) (\mu_{j}-\mu_{l})], \;\;\;j \neq k \neq l.
\label{eqn:b_j}
\end{equation}  

Finally, for  $m=0$, we find $ B_{n0}(t) = 0$ and 
\begin{align}
 A_{n0}(t) &= \sum\limits_{j=1}^{2} c_{j} e^{i \alpha_{j} t},\\
 C_{n0}(t) &= - \frac{1} {f_{1}} \sum\limits_{j=1}^{2} c_{j} (\alpha_{j}  +  V_{11}) e^{i \alpha_{j} t}.
\label{eqn:two_mode_v_c_m0}
\end{align}
Here 
\begin{equation}
 c_{1}  =  \frac{V_{11} + \alpha_{2}} {\alpha_{2}  -  \alpha_{1}},\quad c_{2}  =  \frac{V_{11} + \alpha_{1}} {\alpha_{1}  -  \alpha_{2}}, \\
\label{eqn:two_mode_v_a1}
\end{equation}
and 
\begin{align}
 \alpha_{1} &= {\textstyle\frac{1}{2}} [-y_{1} + (y_{1}^{2} - 4y_{2})^{1/2}], \\
 \alpha_{2} &= {\textstyle\frac{1}{2}} [-y_{1} - (y_{1}^{2} - 4y_{2})^{1/2}],
\label{eqn:two_mode_v_a1}
\end{align}
 where 
\begin{equation}
 y_{1}  =  V_{11}  +  V_{12},\quad y_{2}  =  V_{12} V_{11}  -  f_{1}^{2}.
\label{eqn:two_mode_v_y1}
\end{equation} 

Denoting by $\rho_{1}(t)$ and $\rho_{2}(t)$, the reduced density matrices for the fields $F_{1}$ and  $F_{2}$ respectively, we get (after dropping the explicit time dependence from each of the coefficients)
\begin{widetext}
\begin{align}
 \bra{n} \rho_{1} (t) \ket{n^{\prime}} &= \sum_{l=0}^{\infty} q_{n} q_{n^{\prime}}^{*} \vert r_{l} \vert^{2} A_{n,l} A_{n^{\prime},l}^{*}  + \sum_{l=1}^{\infty}q_{n+1} q_{n^{\prime}+1}^{*} \vert r_{l-1} \vert^{2} B_{n+1,l-1} B_{n^{\prime}+1,l-1}^{*} + \sum_{l=0}^{\infty} q_{n+1} q_{n^{\prime}+1}^{*} \vert r_{l} \vert^{2} C_{n+1,l} C_{n^{\prime}+1,l}^{*}, \\
 \bra{l} \rho_{2} (t) \ket{l^{\prime}} &= \sum\limits_{n=0}^{\infty} \Big[\vert q_{n} \vert^{2} r_{l} r_{l^{\prime}}^{*} A_{n,l} A_{n,l^{\prime}}^{*} + (1-\delta_{l,0}) (1-\delta_{l^{\prime},0})\big(\vert q_{n+1} \vert^{2} r_{l-1} r_{l^{\prime}-1}^{*} B_{n+1,l-1} B_{n+1,l^{\prime}-1}^{*}\big) + \vert q_{n+1} \vert^{2} r_{l} r_{l^{\prime}}^{*} C_{n+1,l} C_{n+1,l^{\prime}}^{*}\Big].
\label{eqn:two_mode_lambda_rho_f2_matrix_elts}
\end{align}
\end{widetext}

\bibliography{reference}

\end{document}